\documentclass[twocolumn,showpacs,preprintnumbers,amsmath,amssymb]{revtex4}
\input psfig.sty
\usepackage{epsf}
\usepackage{graphicx}  
\usepackage{dcolumn}   
\usepackage{bm}        

\newcommand{\be}{\begin{equation}}
\newcommand{\en}{\end{equation}}
\newcommand{\bea}{\begin{eqnarray}}
\newcommand{\ena}{\end{eqnarray}}
\begin{document}

\title{Current lookback time-redshift bounds on dark energy} %

\author{M. A. Dantas$^1$}
\email{aldinez@on.br}

\author{J. S. Alcaniz$^1$}
\email{alcaniz@on.br}

\author{N. Pires$^2$}
\email{npires@dfte.ufrn.br}

\affiliation{$^1$Observat\'orio Nacional, 20921-400 Rio de Janeiro - RJ, Brasil}

\affiliation{$^2$Universidade Federal do Rio Grande do Norte, 59072-970 Natal - RN, Brasil}


\date{\today}


\begin{abstract}

We investigate observational constraints on dark energy models from lookback time (LT) estimates of 32 old passive galaxies distributed over the redshift interval $0.11 \leq z \leq 1.84$. To build up our LT sample we combine the age measurements for these 32 objects with estimates of the total age of the Universe, as obtained from current CMB data. We show that LT data may provide bounds on the cosmological parameters with accuracy competitive with type Ia Supernova methods. In order to break possible degeneracies between models parameters, we also discuss the bounds when our lookback time versus redshift sample is combined with with the recent measurement of the baryonic acoustic oscillation peak  and the derived age of the Universe from current CMB measurements. 

\end{abstract}


\maketitle

\section{INTRODUCTION}

The ratio of the dark energy pressure to its energy density, the so-called equation of state (EoS) parameter, $\omega$, is nowadays one of the most searched numbers in general relativistic cosmology (see, e.g., Ref.~\cite{revde} for recent reviews). This is so for at least two diferent reasons. First, because if one could set $\omega$ to be exactly $-1$, then there would be a great probability of identifying the dark energy with the vacuum state of all existing fields in the Universe, i.e., the cosmological constant ($\Lambda$). Similarly, if a value $\omega \neq -1$ (or a time variation of $\omega$ over the cosmic evolution) is unambiguously found, then one could not only rule out the cosmological constant but also seriously think of the dark pressure responsible for the current cosmic acceleration as the potential energy density associated with a dynamical scalar field $\phi$. The possibility $\omega \neq -1$ still leads to two diferent routes, i.e., a quintessence field if $-1 < \omega < 1/3$~\cite{quint} or a phantom component for $\omega < -1$~\cite{phantom}. Both cases violate the strong energy condition, $\rho + 3p > 0$, but the latter goes further and also violates the null energy condition, i.e., $\rho + p > 0$ \cite{ec}.

Clearly, constraining the value $\omega$ from diferent sets of observational data constitutes an important way to improve our understanding of the actual nature of the dark energy. In this regard, finding new methods or reviving old ones that could directly or indirectly quantify the amount of dark energy present in the Universe, as well as determine its EoS parameter, are important tasks for both theoretical and observational cosmologists. Bounds on $\omega$ have been obtained from observations based on completely distinct physics~\cite{sne,cmb,arcd,ohrg,angs,xray,bao,bao1,gl} (see also \cite{revde} and Refs. therein). This diversity of technics, as well as combinations among them, is particularly important to a more realiable determination of $\omega$, since diferent methods may constrain different regions of the parameter space and, therefore, may be complementary to each other. 

In this paper, by following the methodology presented in Refs.~\cite{lt,mad}, we are particularly interested in deriving
the current lookback time (LT) versus redshift bounds on the dark energy EoS and its density parameter from a sample of 32 passively evolving galaxies, as recently studied by Simon \textit{et al.}~\cite{svj}. The absolute age for these 32 objects was determined by fitting stellar population models and the sample includes observations from the Gemini Deep Deep Survey (GDDS)~\cite{gdds} and archival data~\cite{arcd}. The same data, along with other age estimates of high-$z$ objects, were recently used to reconstruct the shape and redshift evolution of the dark energy potential~\cite{svj}, to place bounds on holography-inspired dark energy scenarios~\cite{china}, as well as to constrain models of modified gravity~\cite{jcap}. Here, we focus our analysis on two dark energy models, namely, the standard $\Lambda$CDM scenario and a flat universe driven by non-relativistic matter (baryonic + dark) and a smooth negative-pressure dark energy component ($\omega$CDM). In order to better constrain the parametric spaces for these scenarios, we also combine LT data with the recent measurement of the baryonic acoustic oscillation peak \cite{bao} and the derived age of the Universe from current CMB measurements~\cite{cmb}.

\begin{figure*}
\label{fig:trans}
\centerline{\psfig{figure=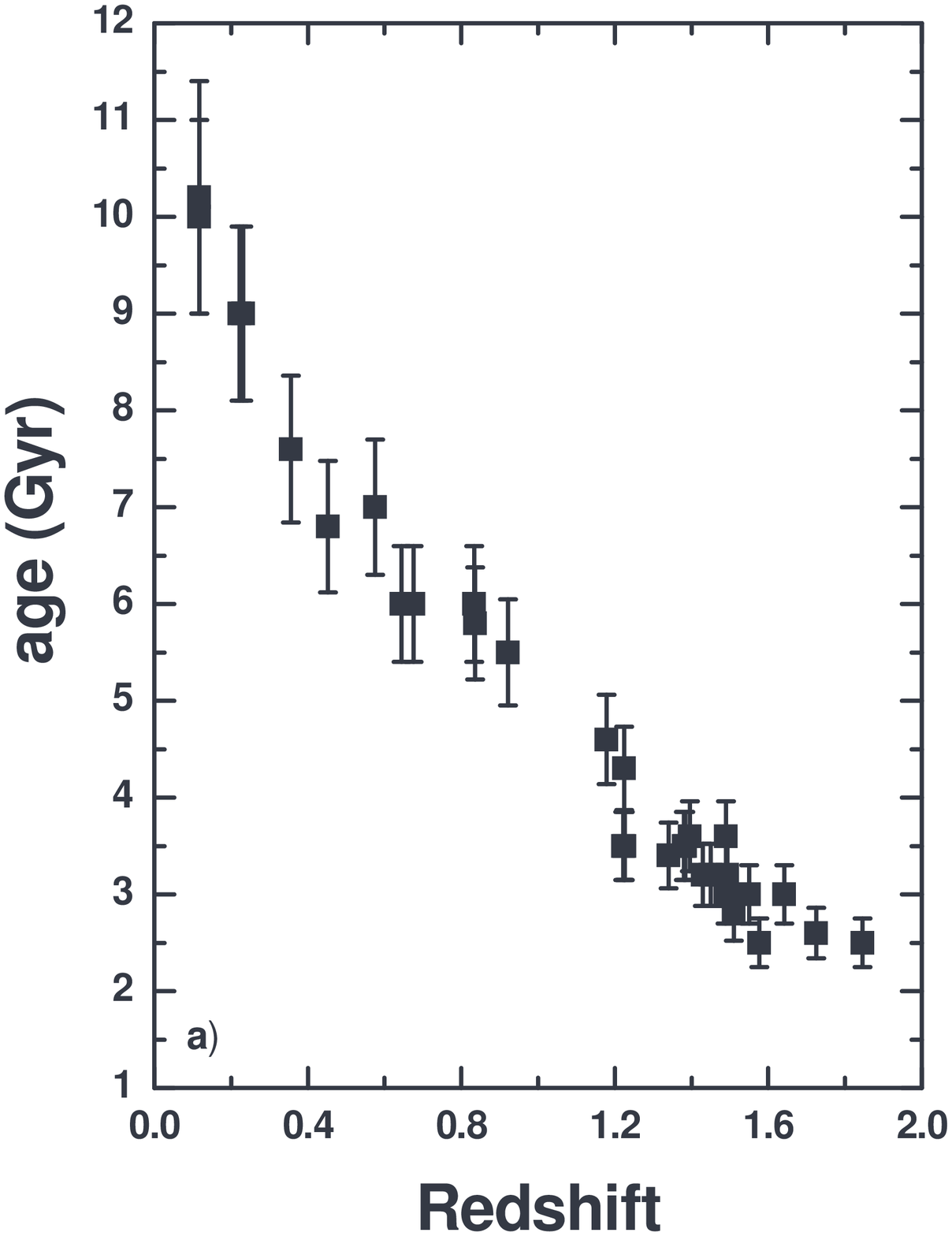,width=3.5truein,height=3.0truein,angle=0}
\psfig{figure=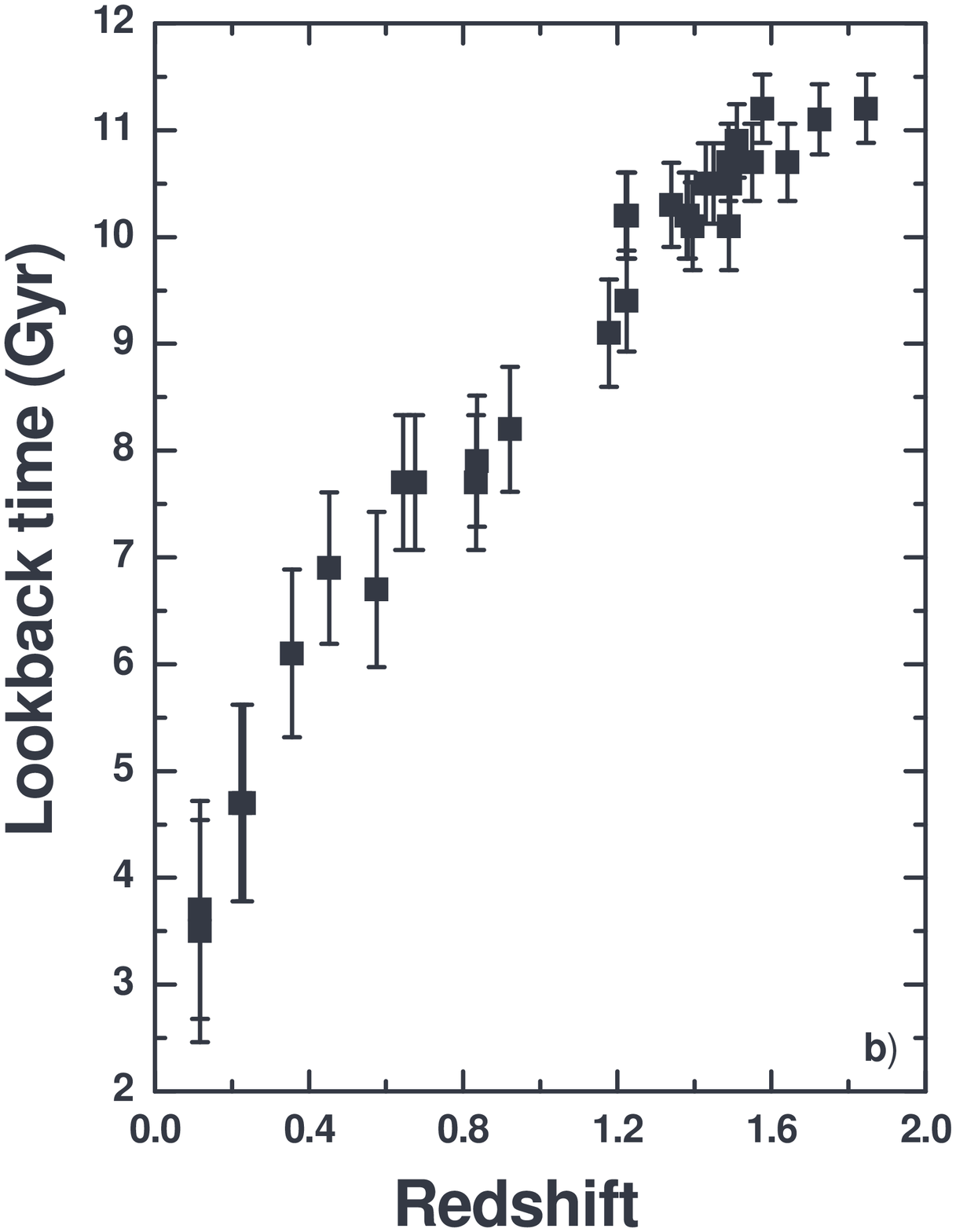,width=3.5truein,height=3.0truein,angle=0}
\hskip 0.1in}
\caption{The age-redshift data points. {\bf Left:} Original data from Ref.~\cite{svj}. This sample corresponds to 32 old passive galaxies distributed over the redshift interval $0.11 \leq z \leq 1.84$ and includes observations from the Gemini Deep Deep Survey (GDDS)~\cite{gdds} and archival data~\cite{arcd}. {\bf Right:} The LT sample. To obtain this sample we have combined the age measurements for these 32 objects with estimates of the total age of the Universe, $t_{0}^{obs} = 13.7 \pm 0.2$ Gyr, as obtained from current CMB data \cite{cmb}.}
\end{figure*}


\section{LOOKBACK TIME-REDSHFT TEST}

\subsection{Theory}
The lookback time-redshift relation, defined as the difference between the present age of the Universe ($t_0$) and its age ($t_z$) when a particular light ray at redshift $z$ was emitted, can be written as
\begin{equation} \label{looktheo}
t_L(z;\mathbf{p}) = {\rm{H}}^{-1}_0 \int_o^z{\frac{dz'}{(1 + z'){{\cal{H}}(\mathbf{p})}}},
\end{equation}
where ${\rm{H}}^{-1}_0 = 9.78h^{-1}$ Gyr and $h$ ranges in the HST \textit{key} project 1$\sigma$ interval $0.64 \leq h \leq 0.8$~\cite{hst}. In the above expression,
 \begin{eqnarray} \label{one}
{\cal{H}}(\mathbf{p})  =  \sqrt{\rm{E}(\mathbf{p})} \quad \rm{with} \quad \rm{E}(\mathbf{p}) = \Omega_m a^{-3} + \Omega_{\textit{x}} a^{-3(1+\omega)}
\end{eqnarray}
where ${\cal{H}}(\mathbf{p}) \equiv \rm{H}(\textbf{p})/\rm{H}_0$ , the complete set of parameters is $\mathbf{p} \equiv  {\Omega_j ,\omega}$ ($j \equiv \rm{m}, \Lambda$ and $x$ stand for matter, cosmological constant ($\omega = -1$) and dark energy ($\omega < 0$) density parameters, respectively), and the subscript $0$ denotes present-day quantities.

To proceed further, let us now consider an object at redshift $z_i$ whose the age $t(z_i)$ is defined as the difference between the age of the Universe at $z_i$ and the one when the object was born $z_{F}$,  i.e.,
\begin{eqnarray} \label{five}
t(z_i)  =  {1 \over {\rm{H_0}}} \left[\int_{z_i}^{\infty}{\frac{dz'}{(1 + z'){{\cal{H}}(\mathbf{p})}}} -  \int_{z_F}^{\infty}{\frac{dz'}{(1 + z'){{\cal{H}}(\mathbf{p})}}}\right]
\end{eqnarray}
or, equivalently,
\begin{equation}  \label{six}
t(z_i)  = t_L(z_F) - t_L(z_i).
\end{equation}
From the above expressions, we can define the observed lookback time to an object at $z_i$ as~\cite{lt}
\begin{eqnarray} \label{lookobs}
 t^{obs}_L(z_i; \tau) & = & t_L(z_F) - t(z_i)  \nonumber \\ & &
= [t^{obs}_o - t(z_i)] - [t^{obs}_o - t_L(z_F)] \nonumber \\ & &
= t^{obs}_o - t(z_i) - \tau_i,    
\end{eqnarray}
where $\tau_i$ stands for the so-colled delay factor, which accounts
for our ignorance about  the amount of time since the beginning of the structure
formation in the Universe until the formation time ($t_f^i$) of the object $i$~\cite{arcd}.

\begin{figure*}
\label{fig:trans}
\centerline{\psfig{figure=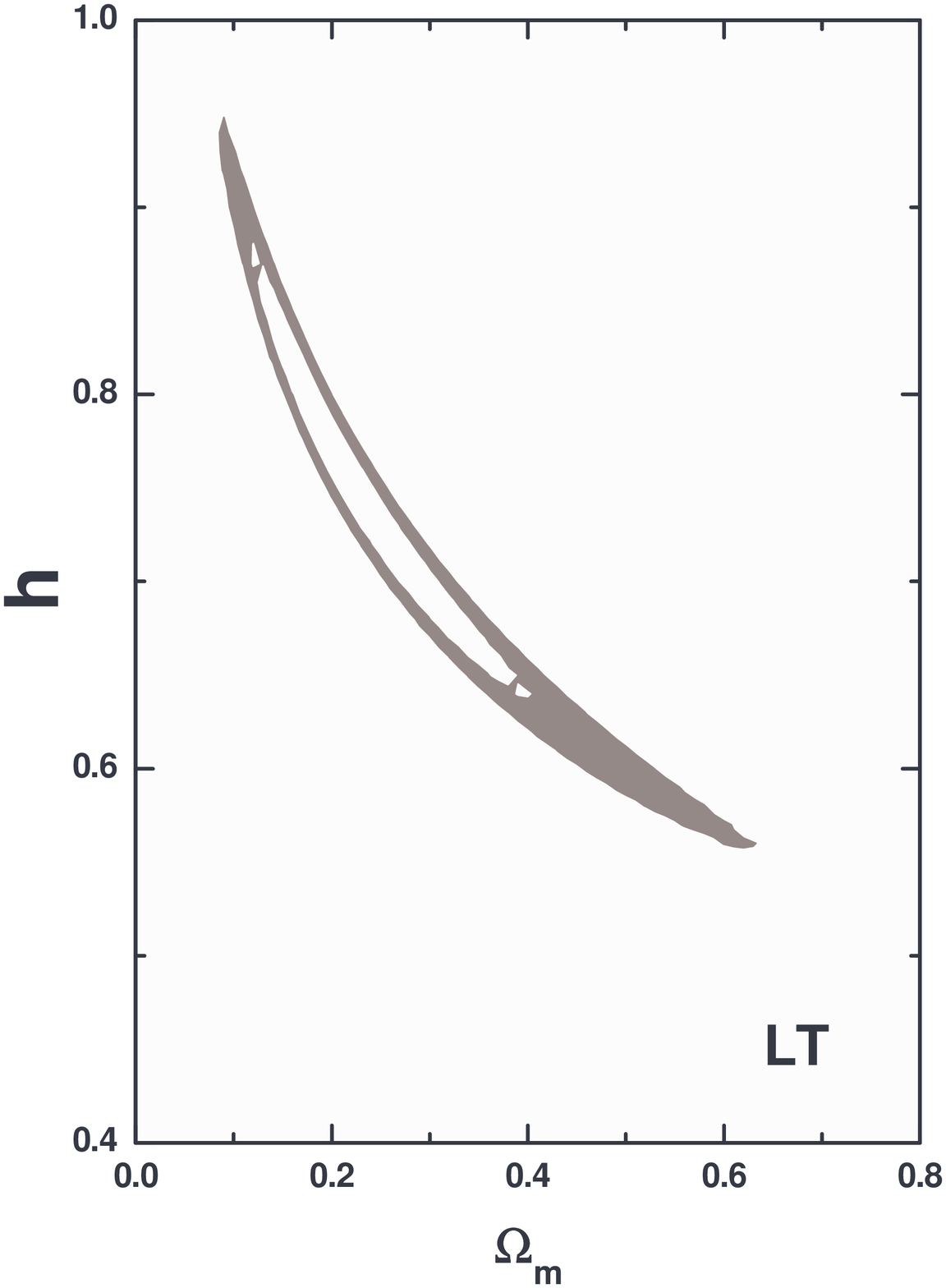,width=2.3truein,height=2.3truein,angle=0}
\psfig{figure=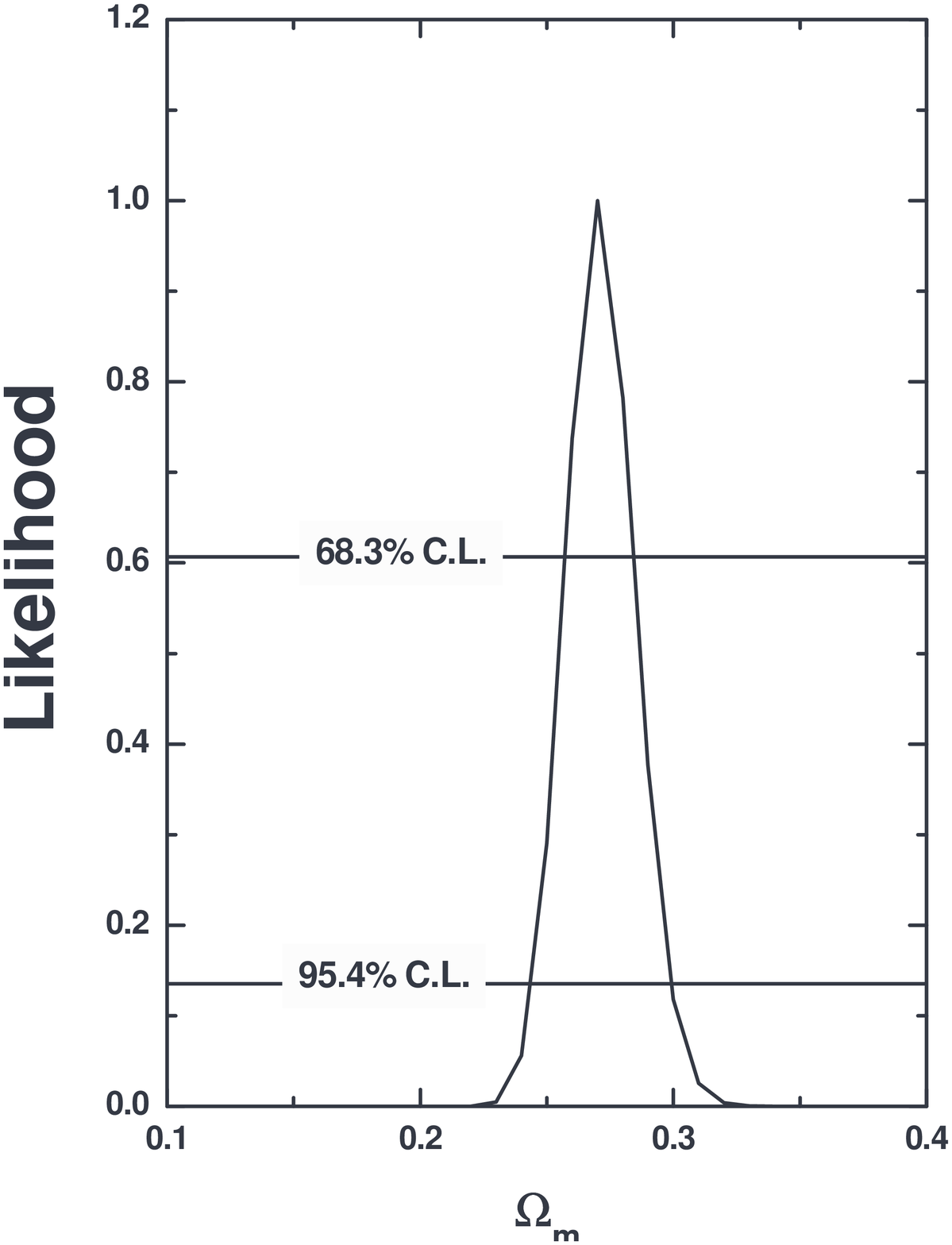,width=2.3truein,height=2.3truein,angle=0}
\psfig{figure=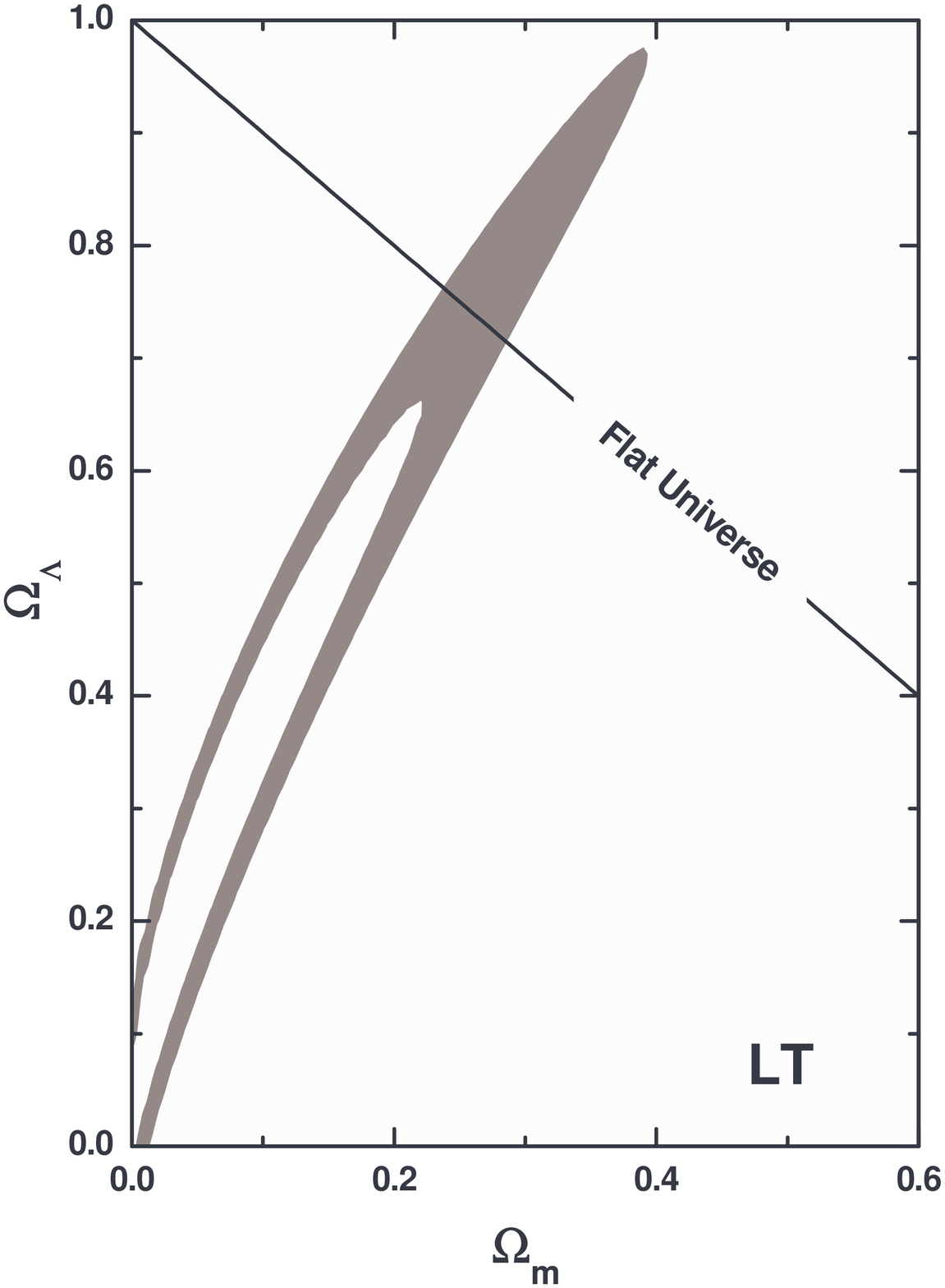,width=2.3truein,height=2.3truein,angle=0}
\hskip 0.1in}
\caption{{\bf Left:} Confidence contours at 68.3\% and 95.4\% in the parametric space $\Omega_{\rm{m}} - h$ from LT analysis. As physically expected, the larger the value of $\Omega_{\rm{m}}$ the smaller the value of the Hubble parameter that is allowed in order to fit the LT data. {\bf Middle:} Likelihood function for $\Omega_{\rm{m}}$. From this calculation, we have found $\Omega_{\rm{m}} = 0.259 \pm 0.030$ at $95.4\%$ (C.L.). {\bf Right:} The $\Omega_{\rm{m}} - \Omega_{\Lambda}$ plane arsing from our LT analysis. The best-fit favours a spatially open universe with $\Omega_k \simeq 0.7$.}
\end{figure*}

We estimate the best-fit to the set of parameters $\mathbf{p}$ by defining the likelihood function
\begin{equation} \label{six}
{\cal{L}}_{age} \propto \exp\left[-\chi_{age}^{2}(z;\mathbf{p},\tau_i)/2\right],
\end{equation}
where $\chi_{age}^{2}$ is given by
\begin{eqnarray} \label{chi2}
\chi_{age}^{2} &  = & \sum_{i=1}^{n}{\frac{\left[t_L(z_i;\mathbf{p}) -
t^{obs}_L(z_i; \tau_i)\right]^{2}} {\tilde{\sigma}_{\rm{i}}^{2}}} + \nonumber \\ & &
\quad \quad \quad \quad   + \frac{\left[t_o(\mathbf{p}) -
t^{obs}_o\right]^{2}}{\sigma_{t^{obs}_o}^{2}}.
\end{eqnarray}
Here, $\tilde{\sigma}_{\rm{i}}^{2} \equiv \sigma_i^{2} + \sigma_{t^{obs}_o}^{2}$, $\sigma_i^{2}$ is the uncertainty in the individual lookback time to the i$^{\rm{th}}$ galaxy of our sample and $\sigma_{t^{obs}_o}$ stands for the uncertainty on the total expanding age of the Universe ($t^{obs}_o$). The prior on the total age of the Universe is justified by the fact that quintessence scenarios that are able to explain age estimates of high-$z$ objects may not be compatible with the total expanding age up to $z = 0$ (and vice-versa) (see, e.g, \cite{friaca} for a discussion). 

Another important aspect concerns the delay factor $\tau$. Note that in principle there must be variations in the value of $\tau$ for each object in the sample (galaxies form at different epochs). Differently from Ref.~\cite{mad}, in the present analysis the delay factor $\tau_i$ for \emph{each object} of the sample is assumed as a nuisance parameter, so that we marginalize over them. Note also that, although involving a $n$ number of integrations, this marginalization may also be analytically obtained by defining a  modified log-likelihood function $\widetilde{\chi}^{\,2}$, i.e.,
\begin{eqnarray} \label{eight}
\widetilde{\chi}^{\,2} & = & -2\ln\left({\int_{0}^{\infty}...\int_{0}^{\infty}d{\tau_i}\exp
\left[-\frac{1}{2}\chi_{age}^{2}(z;\mathbf{p},\tau_i)\right]}\right)\\
& = &  n\ln\left(\frac{2}{\pi}\right) + \sum_{i=1}^{n}{\ln\left(\frac{1}{\tilde{\sigma}_i^2}\right) - 2\sum_{i=1}^{n}{\ln\left[{\rm{erfc}}({\rm{A}}_i)\right]}} + E, \nonumber
\end{eqnarray}
where
\begin{eqnarray}
{\rm{A}}_i = \left(\frac{\Delta_i}{\sqrt{2}\tilde{\sigma}_i}\right),\; \; \; \; \Delta_i = t_L(z_i;\mathbf{p}) - [t^{obs}_o - t(z_i)], \nonumber
\end{eqnarray}
and $E$ is the second term of the rhs of Eq. (\ref{chi2}).



\begin{figure*}[t]
\label{fig:trans}
\centerline{\psfig{figure=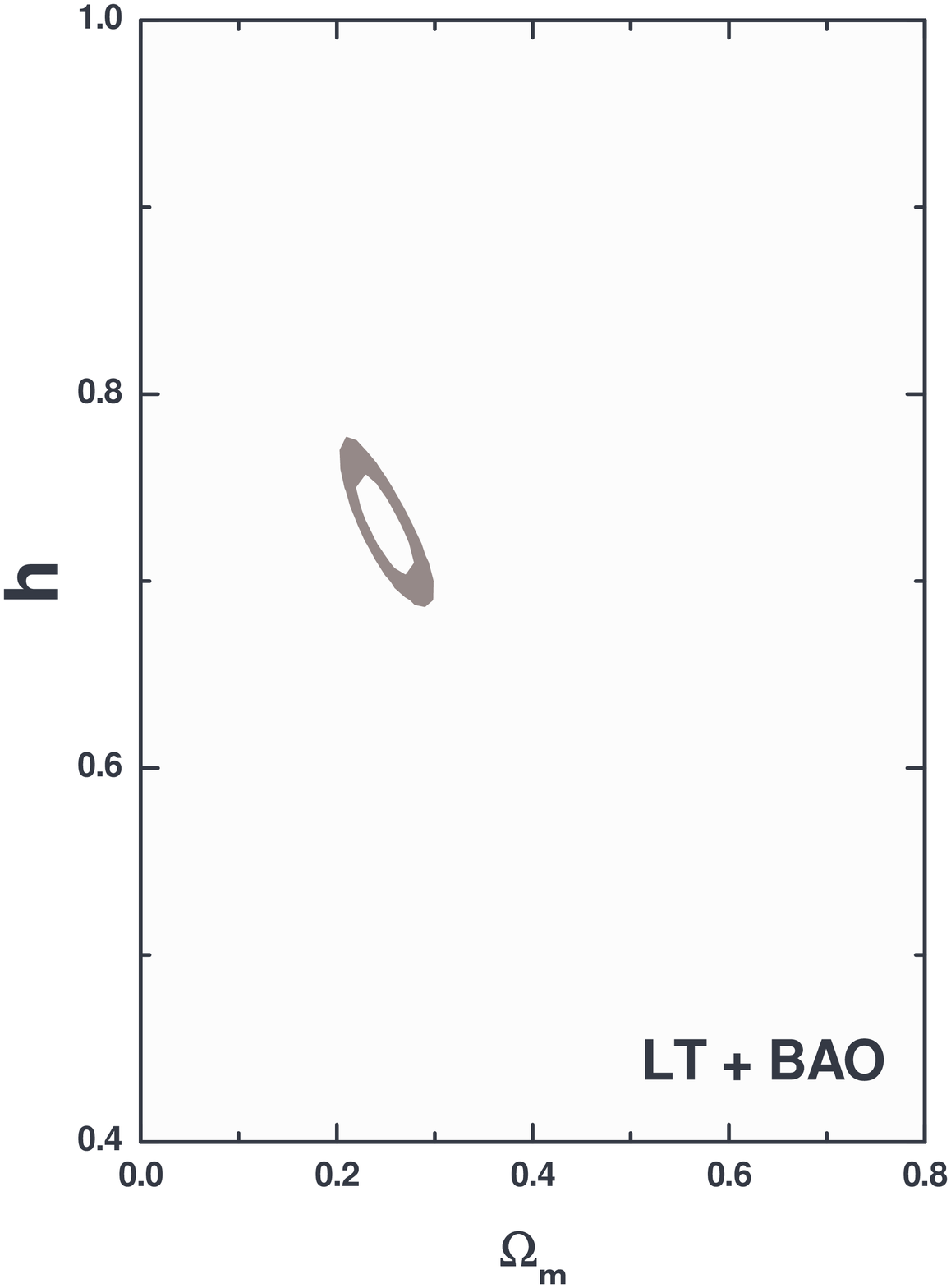,width=3.7truein,height=2.7truein,angle=0}
\psfig{figure=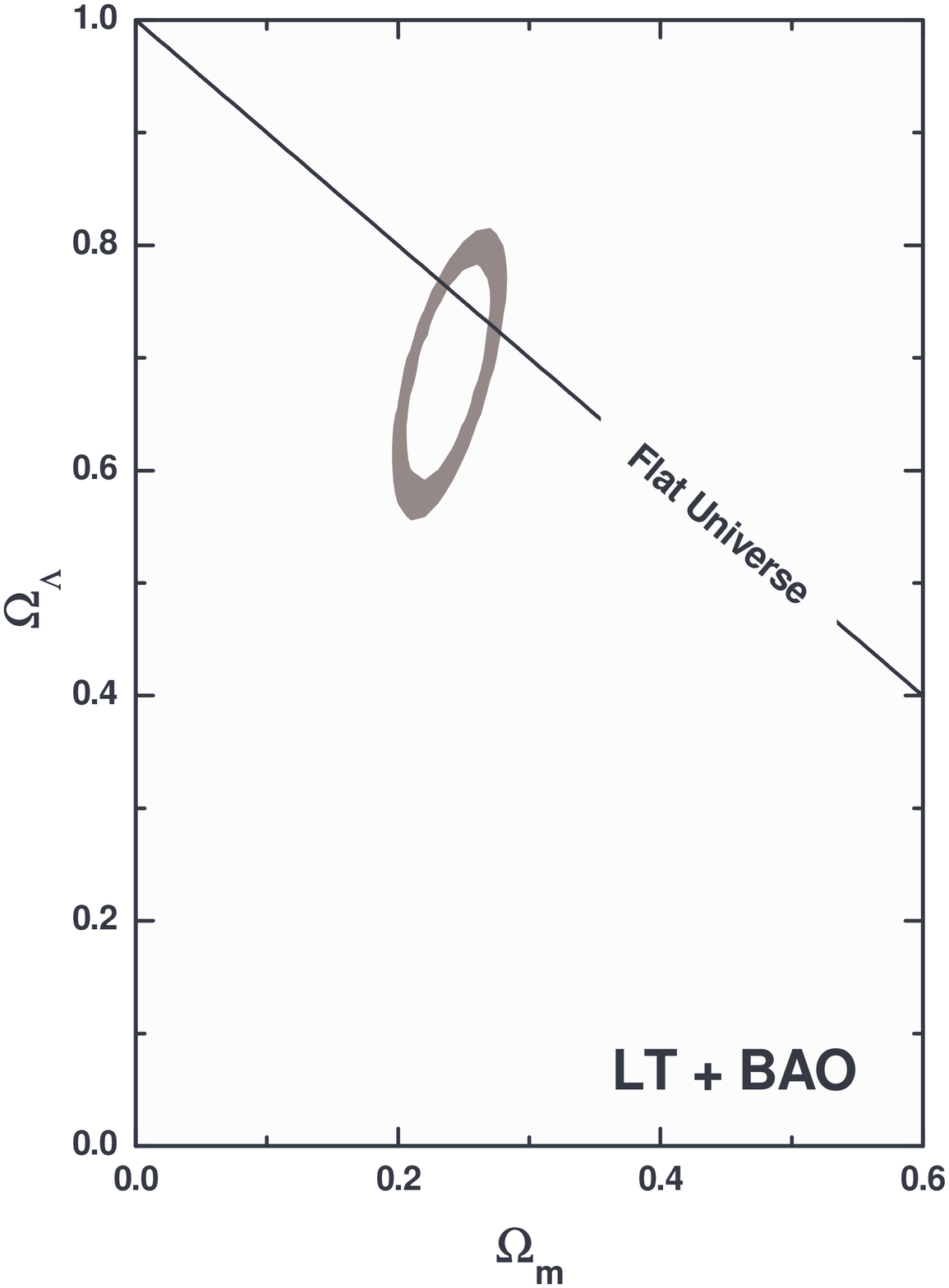,width=3.7truein,height=2.7truein,angle=0}
\hskip 0.1in}
\caption{Results from joint analyses involving LT and BAO data. {\bf Left:} The plane $\Omega_{\rm{m}} - h$ for a flat $\Lambda$CDM model. From this analysis we found $\Omega_{\rm{m}} = 0.25 \pm 0.02$ and $h = 0.73^{+0.02}_{-0.03}$ at $2\sigma$ level. {\bf Left:} 68.3\% and 95.4\% confidence contours in the $\Omega_{\rm{m}} - \Omega_{\Lambda}$. The best-fit model clearly favours a spatially flat model with $\Omega_k  \simeq 0.07$ and $\Omega_{\rm{m}} = 0.24 \pm 0.02$ and $\Omega_\Lambda = 0.69 \pm 0.08$ at $95.4\%$ (C.L.).}
\end{figure*}

\subsection{Data}
In order to apply the method outlined above, we use age estimates of 32 old passive galaxies distributed over the redshift interval $0.11 \leq z \leq 1.84$, as recently analized in Ref.~\cite{svj}. The total sample is composed by three sub-samples: field early-type galaxies from Ref.~\cite{treu}, whose ages were obtained by using SPEED models of Ref.~\cite{speed}; 20 red galaxies from the publicly released Gemini Deep Survey (GDDS)~\cite{gdds} - Ref.~\cite{svj} re-analized the GDDS old sample by using a different stellar population models and obtained ages within 0.1 Gyr of the GDDS collaboration estimates - and the two radio galaxies LBDS 53W091 and LBDS 53W069~\cite{arcd}.

To build up our LT sample, we combine the ages of the above galaxy sample with estimates of the total age of the Universe $t_0^{obs}$, according to Eq. (5). In our analysis, we assume $t_0^{obs} = 13.7 \pm 0.2$ Gyr, as obtained from a joint analysis involving current data of the most recent CMB experiments (WMAP, DASI, VSA, ACBAR, MAXIMA, CBI and BOOMERANG)~\cite{cmb}. In Panels (1a) and (1b) we show, respectively, the original age estimates and transformed lookback time as a function of the redshift for the 32 galaxies of Ref.~\cite{svj}.


\section{Results}

In this Section we discuss quantitatively how our LT sample may place bounds on the EoS and the density parameters describing the dark matter and dark energy through a statistical analysis of the data. We perform our analyses in the context of two different dark energy models, namely, the standard $\Lambda$CDM scenario (flat and general curvature) and a spatially flat universe driven by non-relativistic matter (baryonic + dark) and a negative-pressure dark energy component ($\omega$CDM), as described in Eq. (\ref{one}).

\subsection{$\Lambda$CDM}
In Figures (2a)-(2c) we show the first results of our statistical analyses. By fixing $\omega = -1$ in Eq. (\ref{one}), Panel 2(a) shows contour plots ($68.3\%$ and $95.4\%$ C.L.) in the $\Omega_{\rm{m}} - h$ plane for the $\chi^2_{age}$ given by Eqs. (\ref{six})-(\ref{eight}) plus a Gaussian prior on the Hubble parameter, $h = 0.72\pm0.08$, as given by the final results of the HST \textit{key} project~\cite{hst}. As physically expected, the larger the value of $\Omega_{\rm{m}}$ the smaller the value of the Hubble parameter that is allowed by the statistical analysis in order to fit the $t_L(z)$ estimates. At $95.4\%$ (C.L.), we have found $0.18 \leq \Omega_{\rm{m}} \leq 0.23$ or, equivalently, $0.74 \leq \Omega_\Lambda \leq 0.77 \quad (\Omega_\Lambda = 1 - \Omega_{\rm{m}} )$.

Panel 2(b) shows the likelihood function for the matter density parameter. The dotted lines are cuts in the regions of $68.3\%$ and $95.4\%$ (C.L.). From this calculation, we have found $\Omega_{\rm{m}} = 0.259 \pm 0.030$ at $95.4\%$ (C.L.),
which is in good agreement with current $\Omega_{\rm{m}}$ estimates from CMB~\cite{cmb} and other independent~\cite{wm}  results. We also show in Panel 2(c) the $\Omega_{\rm{m}} - \Omega_\Lambda$ space when the flat condition is relaxed, i.e., by adding the term $\Omega_k a^{-2} = (1 - \Omega_{\rm{m}} - \Omega_\Lambda)a^{-2}$ to the E$(\mathbf{p})$ function of Eq. (\ref{one}). Differently to the results from current SNe Ia data (which prefer a spatially closed universe), the best-fit scenario is an open universe with $\Omega_k \simeq 0.7$ and $0.03 \leq \Omega_{\rm{m}} \leq 0.1$ and $0.15 \leq \Omega_\Lambda \leq 0.32$ at $95\%$ (C.L.).


\subsubsection{Joint Analysis}

As well known, the acoustic peaks in the cosmic microwave background (CMB) anisotropy power spectrum is an efficient way for determining cosmological parameters. Because the acoustic oscillations in the relativistic plasma of the early universe will also be imprinted on to the late-time power spectrum of the non-relativistic matter~\cite{peebles}, the acoustic signatures in the large-scale clustering of galaxies yield additional tests for cosmology. In particular, the characteristic and reasonably sharp length scale measured at a wide range of redshifts provide an estimate of the distance-redshift relation, which is a geometric complement to the usual luminosity-distance from SNe Ia. Using a large spectroscopic sample of 46,748 luminous, red galaxies covering 3816 square degrees out to a redshift of $z = 0.47$ from the Sloan Digital Sky Suvey, Eisenstein et al.~\cite{bao} have successfully found the peaks, described by the ${\cal{A}}$-parameter, i.e.{\footnote{ A direct generalization of Eq. (9) for arbitrary curvatures can be  written as ${\cal{A}}\equiv\frac{\Omega_{\rm{m}}^{1/2}}{{\cal{H}}(z_{\rm{BAO}})^{1/3}} \left[\frac{1}{z_{\rm{BAO}}\sqrt{|\Omega_k|}} {\cal{F}}\left( \sqrt{|\Omega_k|} \Gamma(z_{\rm{BAO}})\right) \right]^{2/3} $, where the function ${\cal{F}}(x)$ is defined as $\sinh(x), x,$ and $\sin(x)$, respectively, for open, flat and closed geometries.}},
\begin{eqnarray}
{\cal{A}} & \equiv & \frac{\Omega_{\rm{m}}^{1/2}}{z_{\rm{BAO}}}\left[z_{\rm{BAO}} \frac{\Gamma^{2}(z_{\rm{BAO}};\mathbf{p})}{{\rm{E}}(z_{\rm{BAO}};\mathbf{p})}\right]^{1/3} \nonumber \\ & &
= 0.469(\frac{n_s}{0.98})^{-0.35} \pm 0.017,
\end{eqnarray}
which can be used to constrain cosmological scenarios that do not have a large contribution of dark energy at early times~\cite{bao1}. In the above expression, $z_{\rm{BAO}} = 0.35$ is the redshift at which the acoustic scale has been measured, $\Gamma(z_{\rm{BAO}}; \mathbf{p}) \equiv \int_0^{z_{\rm{BAO}}} dz/{\cal H}(z_{\rm{BAO}})$ is the dimensionless comoving distance to $z_{\rm{BAO}}$, and we have taken the scalar spectral index $n_s = 0.95$, as given in Ref.~\cite{cmb}.


\begin{figure*}[t]
\label{fig:trans}
\centerline{\psfig{figure=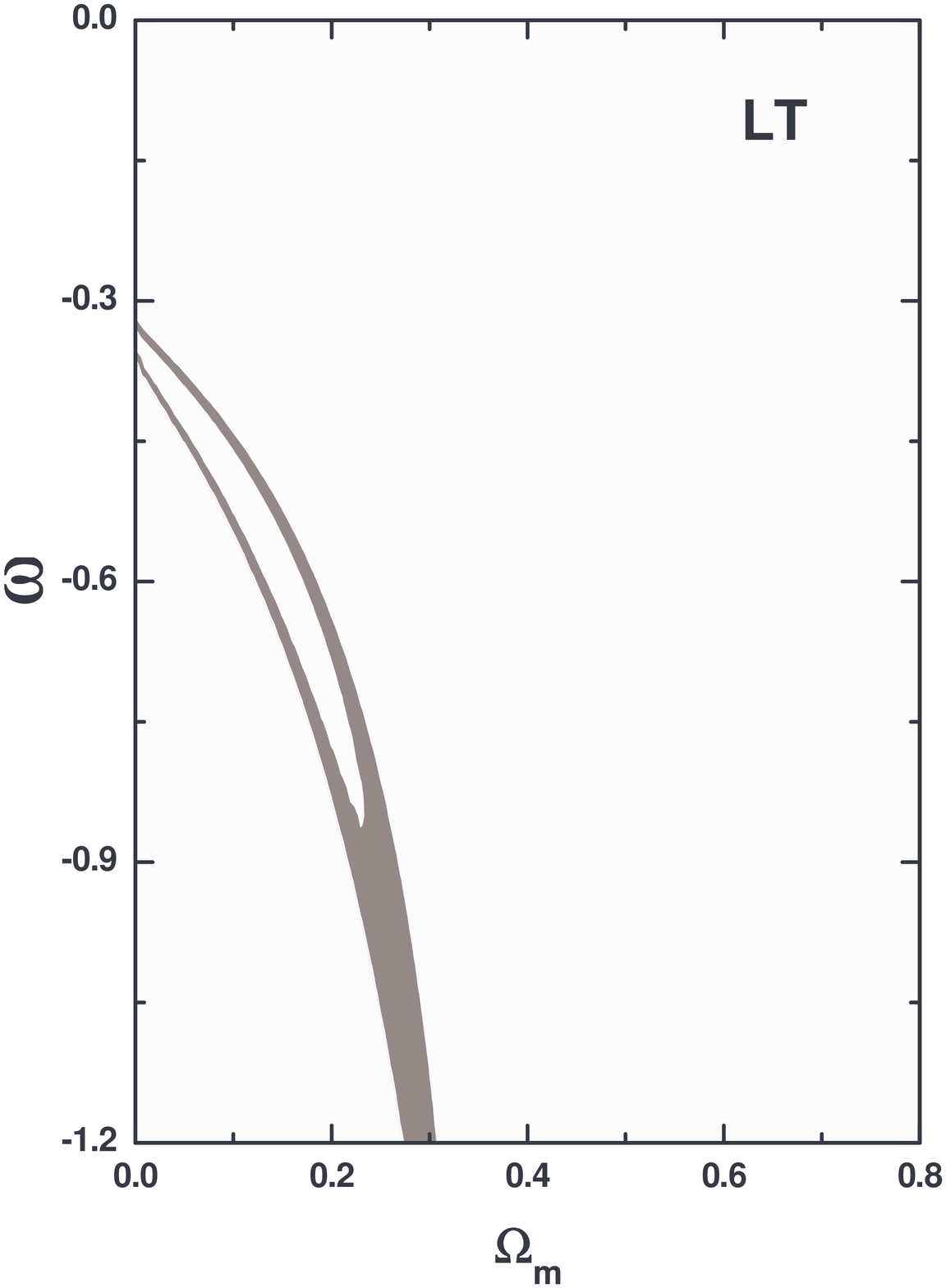,width=3.7truein,height=2.7truein,angle=0}
\psfig{figure=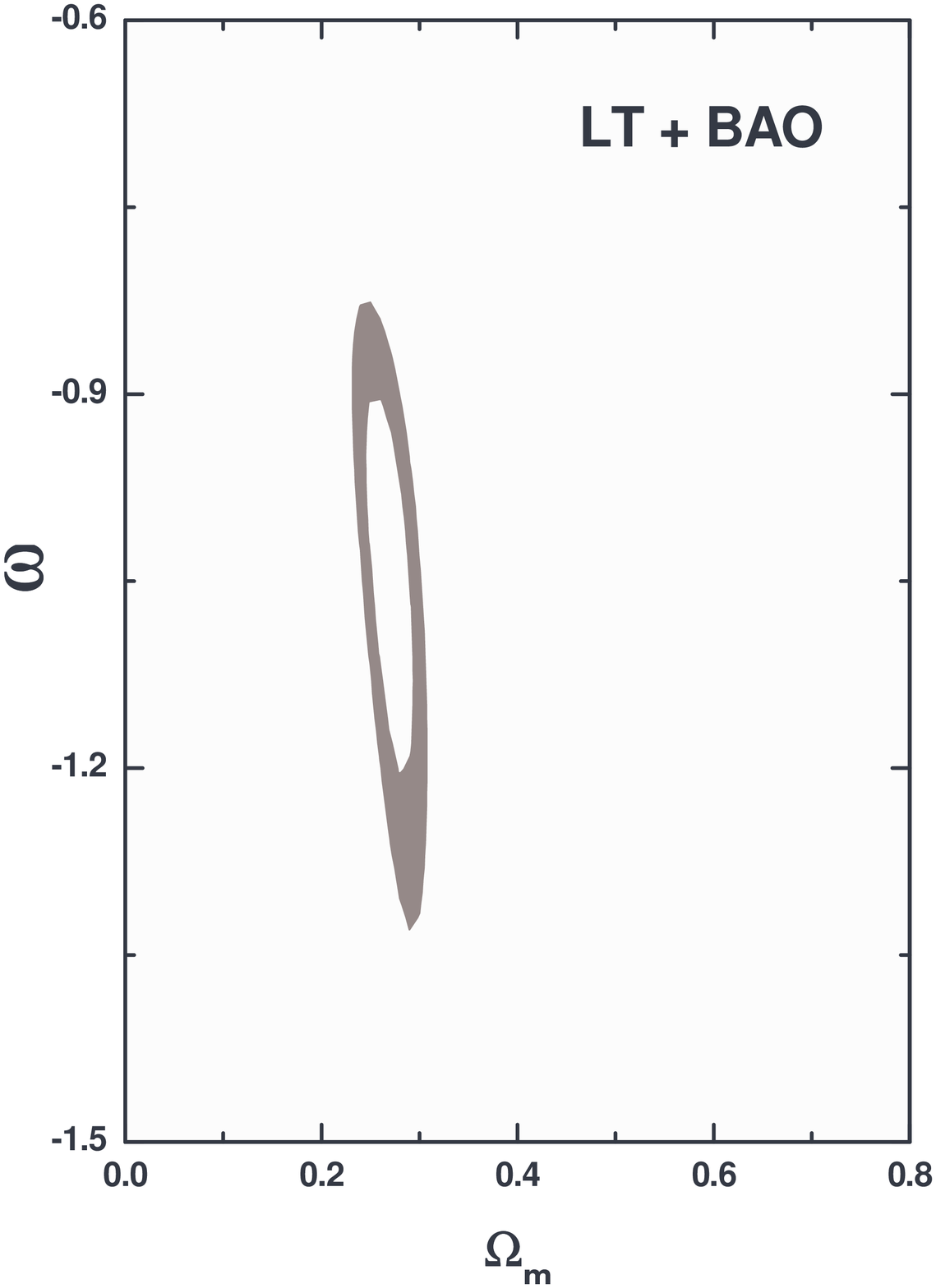,width=3.7truein,height=2.7truein,angle=0}
\hskip 0.1in}
\caption{Results for $\omega$CDM model {\bf Left:} The $\Omega_{\rm{m}} - w$ space allowed by LT data. {\bf Left:} LT + BAO bounds on the $\Omega_{\rm{m}} - \omega$ plane. As discussed in Ref.~\cite{mad}, the tighter results found in this combined analysis reflect the complementarity between LT and BAO measurements.}
\end{figure*}

In Figure (3a) we show contour plots ($68.3\%$ and $95.4\%$ C.L.) in the $\Omega_{\rm{m}}-h$ plane for the LT + BAO combination. Note that, relative to the results shown in Panel (2a), the allowed parameter space is now considerably reduced, with the $2\sigma$ bounds lying in the interval $\Omega_{\rm{m}} = 0.25 \pm 0.02$ and $h = 0.73^{+0.02}_{-0.03}$. By allowing for arbitrary curvature and using a proper generalization of Eq. (9), Fig. (3b) shows the $\Omega_{\rm{m}} - \Omega_\Lambda$ space for the joint LT + BAO analysis. The allowed parameter space now is considerably reduced relative to the former case of Fig. (2c). In agreement with current CMB results~\cite{cmb}, the best-fit model for this joint LT + BAO analysis clearly favors a nearly flat universe with $\Omega_k  \simeq 0.07$ and $\Omega_{\rm{m}} = 0.24 \pm 0.02$ and $\Omega_\Lambda = 0.69 \pm 0.08$ at $95.4\%$ (C.L.).


\subsection{$\omega$CDM}

With the usual assumption that the effective EoS, $w \simeq  \int{\omega(z) \Omega_{x}(z) dz/ \Omega_{x}(z)}$, is a good approximation for the wide class of dark energy scenarios, from now on we discuss the bounds from LT and LT+BAO data on $\omega$.

Figure (4a) shows the parametric space $\Omega_{\rm{m}} - \omega$ allowed at $68.3\%$ and $95.4\%$ (C.L.) from LT data only.
Note that, although the matter density parameter is well constrained by these data, a large interval for $\omega$ is still allowed. In particular, the best-fit model happens for values of $\Omega_{\rm{m}} \simeq 0.1$ and $\omega  \simeq -0.46$. Note also that, although the physics behind LT and SNe Ia observations are quite different, LT constraints on the $\Omega_{\rm{m}} - \omega$ plane are very similar to those obtained from SNe Ia measurements. This amounts to saying that combinations of LT + SNe Ia data are not able to break possible degeneracies on the plane $\Omega_{\rm{m}} - \omega$, and also that LT data may provide bounds on the cosmological parameters with accuracy competitive with SNe Ia methods (see  Ref.~\cite{mad} for a discussion).

A different result arises when the current BAO measurement at $z = 0.35$ is added to the analysis (Fig. 4b). In this case, both $\omega$ and $\Omega_{\rm{m}}$ intervals are more tightly constrained, with the best-fit values given by $\Omega_{\rm{m}}  \simeq 0.27$ and $\omega \simeq -1.04$. At $95.4\%$ (C.L.) we also found $0.25 \leq \Omega_{\rm{m}} \leq 0.29$ and $-1.21 \leq \omega \leq -0.88$. As discussed in Ref.~\cite{mad}, the tighter results found in the combined analyses [Figs. (3b) and (4b)] reflect the complementarity between LT and BAO measurements, which in turn makes possible to break the degeneracies inherent to the parametric plane $\Omega_{\rm{m}} - \omega$.


\section{FINAL REMARKS}

The recent accumulation of independent observational results has opened up a robust window for probing the behavior of the dark component responsible for the current cosmic acceleration. However, most of the methods employed to place limits on the dark energy EoS ($\omega$) or, more generically, on the parametric space $\Omega_m - \omega$, are essentially based on distance measurements to a particular class of objects or physical rulers (e.g., SNe Ia, CMB, galaxy clusters, etc.). In this regard, it is also particularly important to obtain accurate and independent bounds on the physical behavior of the dark energy, as well as on the other main cosmological parameters, from physics relying on different kinds of observations. 

In this paper, by extending and updating the results of Ref.~\cite{mad}, we have followed this direction and studied
the current constraints on the parametric space $\Omega_m - \omega$ from age measurements of high-$z$ galaxies, as recently discussed in Ref.~\cite{svj}. By using a sample of 32 passively evolving galaxies distributed over the redshift interval $0.11 \leq z \leq 1.84$, we have transformed age measurements into LT estimates (by using the current values for the total age of the Universe from CMB data) and discussed quantitatively how these current age data may constrain the parametric spaces $\Omega_{\rm{m}} - \Omega_{\Lambda}$ and $\Omega_m - \omega$. We have shown that LT data may provide bounds on the cosmological parameters with an accuracy competitive with SNe Ia methods [see, e.g., Figs. (3a) and (3b)]. Due to the complementarity between LT (\textit{age}) and BAO (\textit{distance}) measurements, our best results are obtained when joint analyses involving these two different observables are performed. By assuming $\omega = -1$ ($\Lambda$CDM) and allowing for arbitrary curvature, we have found $\Omega_{\rm{m}} = 0.24 \pm 0.02$ and $\Omega_{\Lambda} = 0.69 \pm 0.08$ at 95.4\% (C.L.), which clearly favours a nearly flat universe with $\Omega_k \simeq 0.07$. For a spatially flat model dominated by a negative-pressure component with a constant EoS $\omega$, we have found $0.25 \lesssim \Omega_{\rm{m}} \lesssim 0.29$ and $-1.21 \lesssim \omega \lesssim -0.88$ (95.4\% C.L.), which is again  close to the so-called \textit{concordance} scenario obtained from the usual distance-based combination of SNe + BAO + CMB data.

\acknowledgments

The authors are very grateful to Deepak Jain, Abha Dev, Simone Daflon and Claudio Bastos for valuable discussions. M.A.D., J.S.A. and N.P. thank CNPq (Brazil) for the grants under which this work was carried out.

\end{document}